%% file: livoxxx.tex
\documentstyle[epsf]{europhys} 

\input{euromacr}

\begin{document}
 
\euro{?}{?}{1-$\infty$}{1999} 
\Date{? ? 1999}
 
\shorttitle{V.\ EYERT \etal ELECTRONIC STRUCTURE OF $ {\rm LIV_2O_4} $}
 
\title{Electronic structure and magnetic interactions in 
       $ {\rm \bf LiV_2O_4} $}
 
\author{V.\ Eyert\inst{1,2}, K.-H.\ H\"ock\inst{2}, S.\ Horn\inst{2}, 
        A.\ Loidl\inst{2} and P.\ S.\ Riseborough\inst{3}}
\institute{\inst{1} Hahn-Meitner-Institut, Glienicker Stra\ss e 100, 
                    D-14109 Berlin \\
           \inst{2} Institut f\"ur Physik, Universit\"at Augsburg, 
                    Universit\"atsstra\ss e 1, D-86135 Augsburg \\ 
           \inst{3} Physics Dept., Polytechnic University, 
                    6 Metro-Tech Center, New York, N.Y.\ 11201} 

\rec{22 October 1998}{in final form 1 April 1999} 

\pacs{
\Pacs{71}{20.Be}{Transition metals and alloys} 
\Pacs{71}{27.+a}{Strongly correlated electron systems; heavy fermions} 
\Pacs{75}{20.Hr}{Local moments in compounds and alloys} 
     }

\maketitle

\begin{abstract}
We present results of all-electron electronic structure calculations for the 
recently discovered $ d $ electron heavy fermion compound $ {\rm LiV_2O_4} $. 
The augmented spherical wave calculations are based on density functional 
theory within the local density approximation. The electronic properties 
near the Fermi energy originate almost exclusively from V $ 3d $ $ t_{2g} $ 
states, which fall into two equally occupied subbands: While $ \sigma $-type 
metal-metal bonding leads to rather broad bands, small $ \pi $-type 
$ p $--$ d $ overlap causes a narrow peak at $ {\rm E_F} $. 
Without the geometric frustration inherent in the crystal structure, 
spin-polarized calculations reveal an antiferromagnetic ground state and 
ferromagnetic order at slightly higher energy. Since direct $ d $--$ d $ 
exchange interaction plays only a minor role, ordering of the localized 
vanadium moments can be attributed exclusively to a rather weak superexchange 
interaction. With the magnetic order suppressed by the geometric frustration, 
the remaining spin fluctuations suggest an explanation of the low temperature 
behaviour of the specific heat. 
\end{abstract}


The magnetic porperties of the metallic spinel structure transition metal 
oxide $ {\rm LiV_2O_4} $ have remained an unresolved issue for some time 
\cite{rogers67,reuter67,kessler71,ueda97}. Apart from a temperature 
independent contribution, the susceptibility was found to follow a Curie-Weiss 
law due to local V magnetic moments \cite{kessler71,ueda97}. The negative 
Weiss temperature of -63 K indicates antiferromagnetic exchange 
interactions between the vanadium spins \cite{kessler71}. However, a 
transition to long-range magnetic order above 1.8 K could be excluded from 
both susceptibility and $ {\rm ^{7}Li} $ NMR measurements 
\cite{ueda97,kondo97}. This apparent contradiction was attributed to possible 
geometric frustration inherent in the vanadium sublattice of the normal spinel  
structure with only nearest neighbour interaction, which might prohibit 
long-range antiferromagnetic order \cite{ueda97,anderson56}.

Recently, Kondo {\em et al.}\ reported a crossover from localized moment to 
heavy Fermi liquid behaviour with a Kondo temperature of $ {\rm T_K \approx} $ 
28 K and an electronic specific heat coefficient of 
$ {\rm \gamma \approx 0.42 J/molK^2} $ at 1 K \cite{kondo97,johnston99}. 
Thus for the first time heavy fermion (HF) behaviour, characteristic of $ f $ 
electron systems, has been observed in a $ d $ electron 
material. Below 30 K, the resistivity exhibited a pronounced smooth downturn 
consistent with Kondo lattice behaviour \cite{kondo97}. The nearly temperature 
independent susceptibility and the $ {\rm ^{7}Li} $ Knight shift in the low 
temperature range \cite{ueda97,kondo97,kondo99} were interpreted as due to 
the disappearance of the V local moments. Neither static magnetic order nor 
superconductivity was observed above 0.02 K \cite{kondo97}. Yet, low 
temperature $ {\rm \mu} $SR data displayed anomalies indicative of static spin 
freezing below 0.8 K \cite{kondo97,merrin98}. The specific heat shows an 
upturn between 0.8 and 0.5 K, which might point to a $ T^3 \ln T $ dependence 
characteristic of large spin fluctuations \cite{doniach66+brinkman68}. Low 
temperature $ {\rm ^{7}Li} $ and $ {\rm ^{51}V} $ NMR data likewise showed 
deviations from uniform heavy Fermi liquid behaviour 
\cite{fujiwara97,fujiwara98,mahajan98}. From neutron diffraction 
experiments, structural transitions between 4 and 295 K could be excluded 
\cite{chmaissem97}. In addition, these measurements revealed a large electronic 
Gr\"uneisen parameter comparable to that of $ f $ electron HF compounds, 
which indicates strong coupling of the electronic and lattice degrees of 
freedom \cite{chmaissem97,johnston99}. In contrast to the $ f $ electron 
systems, however, where HF behaviour as well as long-range antiferromagnetic 
order result from the hybridization of localized and bandlike orbitals, the 
origin of the electron-lattice coupling and the mechanisms leading to HF 
formation in the $ d $ band metal $ {\rm LiV_2O_4} $ are as yet unclear 
\cite{chmaissem97}. 

In this Letter, we present for the first time results of all-electron first 
principles calculations for $ {\rm LiV_2O_4} $. Our extensive study is  
based on density functional theory (DFT) within the local density 
approximation (LDA) \cite{hks}. As a calculational scheme we used the 
scalar-relativistic augmented spherical wave (ASW) method \cite{wkg}. 
As a result, we find (i) a superposition of narrow and broad bands at the 
Fermi energy. Allowing for magnetic ordering of the vanadium moments, we 
obtain (ii) an antiferromagnetic ground state resulting from weak superexchange 
coupling, which is accompanied by ferromagnetic order at a slightly elevated 
energy. Due to the geometric frustration only large spin fluctuations remain, 
while the reduced overlap of the magnetic orbitals points to a possible 
localization transition.


$ {\rm LiV_2O_4} $ crystallizes in the normal spinel structure, which is  
based on an fcc lattice with space group $ Fd\bar{3}m $ ($ O_h^{7} $) 
\cite{anderson56,reuter6066}. We used the values of a = 8.22672 {\AA} for 
the lattice constant and $ {\rm x_O = 0.26111} $ for the oxygen parameter 
as resulting from the 4 K neutron diffraction data \cite{chmaissem97}. 
In the normal spinel structure, the oxygen atoms form an almost perfect cubic 
close-packed sublattice with tetrahedral and octahedral interstices. The 
lithium atoms occupy 1/8 of the tetrahedral sites whereas the vanadium atoms 
occupy half of the octahedral sites and form two interpenetrating fcc 
sublattices of corner sharing tetrahedra with a V--V bond length of 2.9 {\AA} 
\cite{anderson56}. Due to cation ordering the crystal field at the transition 
metal sites is trigonal rather than cubic. The deviation of the oxygen 
$ {\rm x_O} $ parameter from the ideal value 0.25 causes a shift of 
0.158 {\AA} in $ \langle111\rangle $ direction, which enlarges/compresses 
the available volume at the tetrahedral/octahedral sites; the V--O distance 
is decreased by 4.6\%. At the same time, the V--O--V bond angle changes 
to $ 95.2^{\circ} $.

\begin{figure}[htb]
\centerline{\hbox{
\epsfxsize=10.0cm \epsfbox{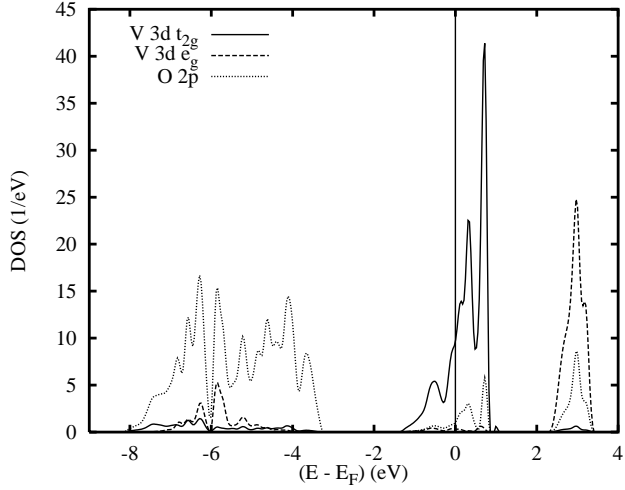}
}}
\caption{Partial densities of states (DOS) of $ {\rm LiV_2O_4} $ per 
         unit cell.} 
\label{fig:sect41}   
\end{figure}

Fig.\ \ref{fig:sect41} 
displays the dominant partial densities of states (DOS). States not 
included play only a negligible role in the energy interval shown. Three 
groups of bands are identified: While bands in the energy range from 
-8 to -3.3 eV originate mainly from O $ 2p $ states and have only a small  
admixture from V $ 3d $ states, the upper two groups of bands, which extend 
from -1.3 to 0.8 eV and 2.4 to 3.4 eV, are predominantly derived from the 
V $ 3d $ states. Although additional O $ 2p $ contributions are apparent in 
this energy range, $ p $--$ d $ hybridization is much reduced as compared 
to other early transition metal oxides \cite{vpop}. As concerns the occupied 
states our findings are in good agreement with the photoemission data 
\cite{fujimori88}. 
Octahedral crystal field splitting, as expected from the nearly cubic 
coordination of the V atoms by the oxygen atoms, leads to clear energetical 
separation of the $ 3d $ $ t_{2g} $ and $ e_g $ groups of bands as is 
visible in fig.\ \ref{fig:sect41}. Whereas the former states appear 
exclusively around $ {\rm E_F} $, the $ e_g $ states dominate at higher 
energies. Contributions of the V $ 3d $ states to the oxygen derived bands 
originate almost exclusively from the $ e_g $ states, which, forming 
$ \sigma $ bonds, experience a larger overlap with the O $ 2p $ states. 
In contrast, the $ t_{2g} $ orbitals, which give rise to $ \pi $ bonds, 
yield only a negligible contribution in this energy range. In addition to 
these rather weak $ p $--$ d $ bonds the $ t_{2g} $ states experience, 
however, strong $ \sigma $-type overlap with the $ t_{2g} $ orbitals at 
neighbouring vanadium sites of the fcc sublattice. Hence, these $ d $ states 
take part in two different types of bonding, namely $ \sigma $-type V--V- 
and $ \pi $-type V--O-bonding, which leads to two different band widths. 
Nevertheless, since both the metal-metal and the metal-oxygen bonding are 
mediated by the {\em same} orbitals, a simple analysis of the partial DOS 
would not allow to distinguish the different roles played by the $ t_{2g} $ 
orbitals. 

\begin{figure}[htb]
\centerline{\hbox{
\hspace*{1.0cm}
\epsfxsize=7.0cm \epsfbox{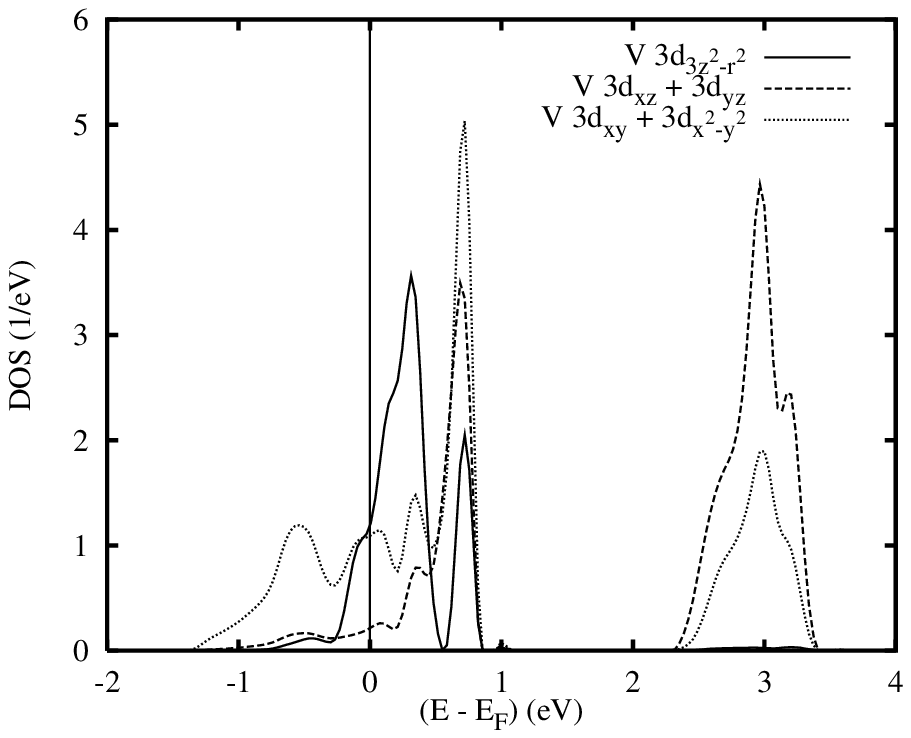}
\epsfxsize=7.0cm \epsfbox{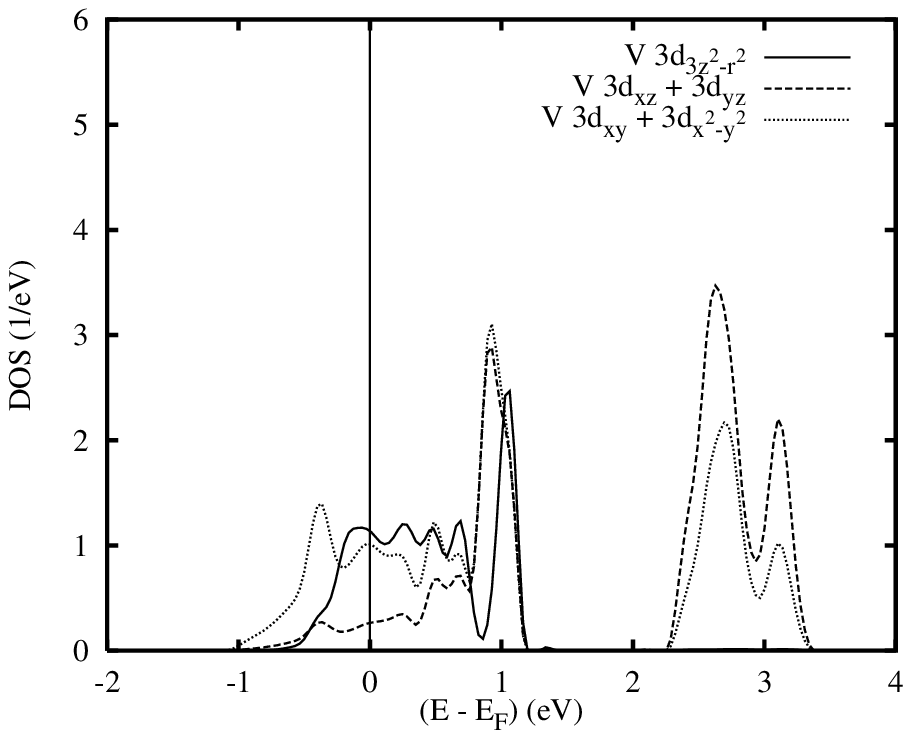}
}}
\caption{Partial densities of states of $ \rm LiV_2O_4 $ for the 
         (a) observed ($ {\rm x_O = 0.26111} $) and 
         (b) ideal ($ {\rm x_O = 0.25} $) structure.} 
\label{fig:sect42}   
\end{figure}

We solved this problem by complementing the calculations for the observed 
crystal structure ($ {\rm x_O = 0.26111} $) with another set, where the 
oxygen positional parameter was hypothetically adjusted to its ideal value 
of $ {\rm x_O = 0.25} $. Since this leaves the vanadium-vanadium bonding 
unaffected, the changes of the electronic structure due to the oxygen 
shift allow to make the aforementioned distinction. The results are shown 
in fig.\ \ref{fig:sect42},  
where we display the partial DOS for the V $ 3d $ dominated bands. We used 
the trigonal representation 
of the fcc lattice and plotted the V $ d_{3z^2-r^2} $, $ d_{xz} + d_{yz} $ 
and $ d_{x^2-y^2} + d_{xy} $ partial DOS. While the $ d_{3z^2-r^2} $ orbitals, 
which correspond to the $ a_{1g} $ orbitals in a trigonal crystal field, point 
along the $ \langle111\rangle $ direction and are of pure $ t_{2g} $ 
character, the other four orbitals comprise a mixture of $ t_{2g} $ and 
$ e_g $ states. However, the latter states are dominated by the 
$ d_{xz} $ and $ d_{yz} $ orbitals, which are almost empty.  

For $ {\rm x_O = 0.26111} $ we observe a clear separation of the 
$ t_{2g} $ states into an $ \approx $2.1 eV wide group formed by the 
$ d_{x^2-y^2} $/$ d_{xy} $ states and another group due to the much narrower 
$ d_{3z^2-r^2} $ bands, which, apart from the peak at 0.7 eV, extend over 
an energy range of only 0.6 eV. On the other hand, for the idealized 
structure this difference of the band widths is removed due to the 
striking change of the $ d_{3z^2-r^2} $ partial DOS: The peak at 
$ \approx $0.35 eV is replaced by a nearly rectangular shaped DOS of 
$ \approx $1.2 eV width. While the O $ 2p $ DOS closely follows these 
changes, the partial DOS due to the other $ d $ orbitals are almost 
unaffected by the oxygen atom shift. From inspection of the band structures 
the drastic changes of the $ d_{3z^2-r^2} $ states can be attributed almost 
exclusively to the striking downshift of a single band by up to 0.45 eV. 
A similar situation has been found for $ {\rm LiTi_2O_4} $ 
\cite{satpathy87,massidda88}. While in the real structure, this band lies 
completely above $ {\rm E_F} $ and causes the peak of the $ d_{3z^2-r^2} $ 
partial DOS at 0.35 eV, it is stabilized in the ideal 
structure by the widening of the triangular octahedral faces, which point 
towards the $ \langle111\rangle $ direction. As a consequence, the electron 
count of these orbitals is slightly increased at the expense of the 
$ d_{x^2-y^2} $ and $ d_{xy} $ electrons. 

In conclusion, from their different response to the oxygen atom shifts we 
are able to clearly identify two groups of $ t_{2g} $ bands: The broad 
$ d_{x^2-y^2} $/$ d_{xy} $ states originate from strong $ \sigma $-type 
V--V-overlap and are thus essentially unaffected by the change of the 
oxygen $ {\rm x_O} $ parameter. In contrast, the much narrower 
$ d_{3z^2-r^2} $ bands are involved in weak $ \pi $-type V--O-overlap and 
depend sensitively on the oxygen position. 

Motivated by the observed Curie-Weiss behaviour and disregarding for the 
time being possible geometric frustration effects in the spinel structure,  
we performed additional spin-polarized calculations. Both ferro- and 
antiferromagnetic order with a modulation of spins parallel to one of the 
cubic axes ($ {\bf q} = (0,0,4\pi/a) $) were allowed for and, again, 
we considered both the observed and the idealized crystal structure. In 
all cases we obtained magnetic order with vanadium magnetic 
moments and total energy lowerings as summarized in table \ref{tab:1}. 
\begin{table}[htb]
\caption{Magnetic energy gain and vanadium magnetic moments.}
\label{tab:1}   
\begin{tabular*}{120mm}{l@{\extracolsep\fill}cccc} 
\hline 
\hline 
                           & \multicolumn{2}{c}{$ {\rm x_O = 0.26111} $} 
                           & \multicolumn{2}{c}{$ {\rm x_O = 0.25} $} \\
\cline{2-3}
\cline{4-5}
                           &  ferro &  af    &  ferro &  af    \\ 
\hline 
$ E_{tot} \left[ {\rm mRyd/f.u.} \right] $ 
                           & -1.431 & -3.655 & -2.701 & -1.170 \\
\hline 
$ m_{\rm V} \left[ \mu_B \right] $           
                           &  0.590 &  0.877 &  1.108 &  0.768 \\
\hline 
\hline 
\end{tabular*} 
\end{table}  
The latter are given relative to the respective non-spinpolarized solution. 
The vanadium magnetic moments result from almost equal contributions from 
the $ d_{x^2-y^2} $/$ d_{xy} $ and $ d_{3z^2-r^2} $ orbitals; moments on 
all other atomic spheres are smaller than 0.025 $ {\rm \mu_B} $. 
The main results are: 
For the antiferromagnetic states the rather small change of the ordered 
moment with respect to shifting the oxygen $ {\rm x_O} $ parameter is 
contrasted with a considerable modification of the magnetic energy gain. 
In contrast, we witness a suppression of both the ferromagnetic moment 
{\em and} energy gain on going from the idealized to the observed structure. 
A detailed analysis revealed that this loss of magnetic moment affects all 
the $ t_{2g} $ states to a similar degree. 

The aforementioned behaviour arises from two different mechanisms: 
(i) The isotropic compression of the $ {\rm VO_6} $ octahedra in the 
observed structure as indicated by the 4.6\% decrease of the V--O bond 
length, causes destabilization of both the ferromagnetic moment and order 
as compared to the idealized structure. Furthermore, since the magnetic 
moments carried by the $ d_{x^2-y^2} $/$ d_{xy} $ and $ d_{3z^2-r^2} $ 
orbitals are affected to a similar degree by the oxygen atom shift, 
we conclude that the polarization of the former orbitals is due to 
Hund's rule coupling rather than direct exchange interaction across the 
$ \sigma $-type V--V bonds. 
(ii) The deviation of the V--O--V bond angle from 
$ 90^{\circ} $ for $ {\rm x_O = 0.26111} $ supports antiferromagnetic 
superexchange interaction of the local vanadium magnetic moments via the 
occupied O $ 2p $ orbitals, while changing their size only little. 
The crossover from ferro- to antiferromagnetic order on increasing 
the oxygen $ {\rm x_O} $ parameter and, in particular, the antiferromagnetic 
ground state of the observed structure thus arise from the combined effect 
of an increased overlap of the magnetic orbitals and a suppression of the 
ferromagnetic local vanadium moment. Still, the importance of ferromagnetic 
correlations has been recently revealed by neutron scattering experiments, 
which suggested to characterize $ {\rm LiV_2O_4} $ as an itinerant $ d $ 
electron system close to weak ferromagnetism \cite{krimmel99}. Similar 
conclusions have been drawn from the anomalous temperature dependence of 
the $ {\rm ^{7}Li} $ NMR relaxation rate \cite{fujiwara97,fujiwara98}. 
Finally, the very small values of the magnetic energy gain as compared to 
other transition metal oxides and the almost linear variation of this energy 
gain with the magnetic moment for the ferromagnetic states result from the 
minor role of a direct $ d $--$ d $ exchange interaction as well as the small 
superexchange coupling due to $ \pi $-type $ p $--$ d $ overlap. 
Thus, without the geometric frustration, we would find an antiferromagnetic 
ground state, in which the well localized vanadium magnetic moments are 
coupled by weak superexchange interaction via the O $ 2p $ states. 

Nevertheless, the geometric frustration inherent in the spinel structure with 
only nearest neighbour interactions does prohibit long-range antiferromagnetic 
order as it leads to an energetical degeneracy of different antiferromagnetic 
configurations. Perhaps, in the states with large band width from the direct 
V--V overlap, the frustration of the magnetic interactions allows large 
amplitude spin fluctuations to occur. This could explain the $ T^3 \ln T $ 
behaviour of the specific heat observed at low temperatures, as well as the 
strong field dependence of the magnetic susceptibility below 30 K 
\cite{lohmann99}. These spin fluctuations can result in large quasi-particle 
mass enhancements and thus mimic a heavy fermion state. Alternatively, in 
the absence of magnetic ordering, the states, in which the small overlap of 
the $ d $ with the O $ 2p $ orbitals occurs, could place the orbitals at the 
brink of a transition to a localization and the local spin fluctuations 
could be responsible for the observed heavy fermion behaviour.


To summarize, the electronic properties of $ {\rm LiV_2O_4} $ as resulting 
from electronic structure calculations are dominated by a combination of 
broad and narrow bands near the Fermi energy, which originate from 
strong metal-metal and rather weak metal-oxygen overlap, respectively. 
While the direct V--V bonding does not contribute to magnetic ordering, 
the antiferromagnetic as well as the slightly unstable ferromagnetic state 
result from weak superexchange interaction. However, longe range magnetic 
order is suppressed by the geometric frustration inherent in the spinel 
structure; hence only large spin fluctuations survive, which provide 
a clue to an understanding of the observed specific heat and susceptibility 
data. The picture emerging from the present calculations is different 
from that usually used for $ f $ electron systems, where the hybridization 
between localized and the partially occupied itinerant states causes local 
spin fluctuations responsible for the HF behaviour, and when magnetic 
magnetic ordering occurs it is usually due to the RKKY interaction.

\stars
We gratefully acknowledge valuable discussions with U.\ Eckern, 
D.\ C.\ Johnston, D.\ I.\ Khomskii, and W.\ E.\ Pickett. This work was 
supported by the Deutsche Forschungsgemeinschaft (Forschergruppe HO 955/2).

%
%

\end{document}

%% file: euromacr.tex
\def\etal{{\hbox{{\tenit\ et al.\/}\tenrm :\ }}}

\def\stars{\bigskip\centerline{***}\medskip}

\newif\ifboo \boofalse
